%% file: paper.tex
\documentclass[letter]{aa}
\usepackage{graphicx}
\usepackage{bm}
\usepackage{txfonts}
\usepackage{natbib}
\usepackage{mathrsfs}
\usepackage[colorlinks,allcolors=blue]{hyperref}
\bibpunct{(}{)}{;}{a}{}{,}
\graphicspath{{./fig/}{./png/}}
\input macros

\begin{document}

\authorrunning{K\"apyl\"a}
\titlerunning{Simulations of entropy rain-driven convection}

   \title{Simulations of entropy rain-driven convection}

   \author{P. J. K\"apyl\"a
          \inst{1}
          }

   \institute{Institute for Solar Physics (KIS),
     Georges-K\"ohler-Allee 401a, 79110 Freiburg, Germany email:
     \href{mailto:pkapyla@leibniz-kis.de}{pkapyla@leibniz-kis.de} }

\date{\today}

\abstract{%
    The paradigm of convection in solar-like stars is questioned based
    on recent solar observations.
   }%
   {The primary aim is to study the effects of surface-driven entropy
     rain on convection zone structure and flows.
   }%
   {Simulations of compressible convection in Cartesian geometry with
     non-uniform surface cooling are used. The cooling profile
     includes localized cool patches that drive deeply penetrating
     plumes. Results are compared with cases with uniform cooling.
   }%
   {Sufficiently strong surface driving leads to strong non-locality
     and a largely subadiabatic convectively mixed layer. In such
     cases the net convective energy transport is done almost solely
     by the downflows. The spatial scale of flows decreases with
     increasing number of cooling patches for the vertical flows
     whereas the horizontal flows still peak at large scales.
   }%
   {To reach the plume-dominated regime with a predominantly
     subadiabatic bulk of the convection zone requires significantly
     more efficient entropy rain than what is realized in simulations
     with uniform cooling. It is plausible that this regime is
     realized in the Sun but that it occurs on scales smaller than
     those resolved currently. Current results show that entropy rain
     can lead to largely mildly subadiabatic convection zone, whereas
     its effects for the scale of convection are more subtle.
   }%

  \keywords{convection -- turbulence -- Sun:interior
  }
  \maketitle

\section{Introduction}
\label{sec:intro}

It has become evident during the last decade and a half that the
theoretical understanding of convection in the Sun is not as complete
as previously thought
\citep[e.g.][]{Hanasoge_et_al_2010_ApJ_712_98,Hanasoge_et_al_2012_PNAS_109_11928,Schumacher_Sreenivasan_2020_RvMP_92_041001}. This
is because helioseismology and other solar observations suggest that
velocity amplitudes on large scales on the Sun are much lower than in
current simulations \citep[e.g.][and references
  therein]{Birch_et_al_2024_PhFl_36_117136}. This discrepancy is
commonly referred to as the convective conundrum
\citep{OMara_et_al_2016_AdSpR_58_1475}, and it is the most likely
reason why global magnetohydrodynamic simulations of the Sun struggle
to reproduce the solar differential rotation and large-scale dynamo
\citep[e.g.][and references therein]{Kapyla_et_al_2023_SSRv_219_58}.
This has led to a reevaluation of the fundamental theory of stellar
convection.

In the widely used mixing length theory
\citep[e.g.][]{Vitense_1953_ZAst_32_135,Joyce_Tayar_2023_Galaxies_11_75},
convection is driven locally by an unstable temperature gradient in
accordance with the Schwarzschild criterion.  On the other hand, solar
surface simulations \citep[e.g.][]{Stein_Nordlund_1998_ApJ_499_914}
suggest that convection in the Sun is driven by surface cooling due to
a rain of low entropy material that forms strong downflow plumes. Such
entropy rain is highly non-local and has been suggested to be relevant
also for deep convection \citep{Spruit_1997_MEMSAI_68_397}, and to
lead to a convection zone that is weakly stably stratified
\citep[][]{Brandenburg_2016_ApJ_832_88}. Recent studies of global
Rossby waves in the Sun indeed suggest that the solar convection zone
has to be either very nearly adiabatic
\citep{Gizon_et_al_2021_AA_652_6} or mildly subadiabatic
\citep{Bekki_2024_AA_682_39}.

Simulations of compressible convection have confirmed the existence of
subadiabatic layers in the deep parts of convective envelopes
\citep[e.g.][]{Tremblay_et_al_2015_ApJ_799_142,Bekki_et_al_2017_ApJ_851_74,Hotta_2017_ApJ_843_52,Kapyla_et_al_2017_ApJL_845_23,Brun_et_al_2022_ApJ_926_21,Kapyla_2021_AA_655_78,Kapyla_2024_AA_683_221}. However,
in all of these simulations the subadiabatic layer at the base of the
convection zone encompasses relatively small fraction of the total
depth of the convective layer. It is possible that this is the
configuration in the Sun and other stars with convective
envelopes. Another possibility is that the current simulations do not
capture the entropy rain originating near the surface accurately due
to insufficient resolution. This is plausible because the parameter
regime of the Sun is prohibitively far away from the range accessible
to current simulations
\citep[e.g.][]{Kupka_Muthsam_2017_LRCA_3_1,Kapyla_et_al_2023_SSRv_219_58}. The
resolution issue is particularly dire in the photosphere where the gas
becomes optically thin abruptly over a distance of some tens of
kilometers and coincides with the layer where entropy rain is thought
to be launched. A few earlier numerical studies have explored surface
effects on deep convection zone structure. For example,
\cite{Cossette_Rast_2016_ApJ_829_L16} and
\cite{Yokoi_et_al_2022_MNRAS_516_2718} found increased surface
forcing, corresponding to a steeper entropy gradient, led to reduced
scale of convection in deep layers. A similar approach was used in
\cite{Kapyla_2024_IAUS_365_5} where surface forcing resulted in deeper
convection zones in general due to a thermodynamics-dependent heat
conductivity. On the other hand, \cite{Hotta_Kusano_2021_NatAs_5_1100}
found a negligible effect of the surface layers in simulations
covering the entire density contrast of the solar convection zone.

In the current study the working hypothesis is that the surface
effects occur on scales that cannot be resolved directly in current
simulations. Here the surface-induced entropy rain is modeled by
non-uniform cooling where localized cool patches lead to plume
formation. This is motivated by the non-uniform surface temperature in
the Sun. A similar approach was taken by
\cite{Nelson_et_al_2018_ApJ_859_117} who worked in rotating spherical
shells and included a prescribed impulse to the generated plumes with
the aim to study global phenomena such as differential rotation. At
the other end of the spectrum are the studies of
\cite{Rast_1998_JFM_369_125} and \cite{Anders_et_al_2019_ApJ_884_65}
who studied individual plumes in a non-convecting background. Here
these studies are generalized to cases with a convective background
while retaining simple geometry and omitting the effects of rotation
and magnetism. The primary aim is to explore the transition to entropy
rain-dominated convection and its implications for thermal and flow
structures in the convection zone.

\section{The model}
\label{sec:model}

The model is based on that used in
\cite{Kapyla_2019_AA_631_122,Kapyla_2021_AA_655_78,Kapyla_2024_AA_683_221},
and the {\sc Pencil Code}
\citep[][]{Pencil_Code_Collaboration_2021_JOSS_6_2807}\footnote{\href{https://pencil-code.org/}{https://pencil-code.org}}
was used to produce the simulations. The equations for compressible
hydrodynamics were solved:
\begin{eqnarray}
\frac{D \ln \rho}{D t} &=& -\bm\nabla \bm\cdot \uuu, \label{equ:dens}\\
\frac{D\uuu}{D t} &=& {\bm g} -\frac{1}{\rho}(\bm\nabla p - \bm\nabla \bm\cdot 2 \nu \rho \bm{\mathsf{S}}),\label{equ:mom} \\
T \frac{D s}{D t} &=& -\frac{1}{\rho} \left[\bm\nabla \bm\cdot \left(\FFF_{\rm rad} + \FFF_{\rm SGS}\right) - \mathcal{C} \right] + 2 \nu \bm{\mathsf{S}}^2,
\label{equ:ent}
\end{eqnarray}
where $D/Dt=\pd/\pd t + \uuu\bm\cdot\bm\nabla$ is the advective
derivative, $\rho$ is the density, $\uuu$ is the velocity,
$\bm{g}\!=\!-g\hat{\bm{e}}_z$ is the acceleration due to gravity, $p$
is the pressure, $T$ is the temperature, $s$ is the specific entropy,
and $\nu$ is the constant kinematic viscosity. The terms $\FFF_{\rm
  rad}$ and $\FFF_{\rm SGS}$ are the radiative and turbulent
subgrid-scale (SGS) fluxes, respectively, and $\mathcal{C}$ describes
cooling near the surface. $\SSt$ is the traceless rate-of-strain
tensor with $\SStij=\onehalf (u_{i,j}+u_{j,i})-\onethird \delta_{ij}
\bm\nabla\bm\cdot\uuu$. The gas is assumed to be optically thick and
fully ionized where radiation is modeled via the diffusion
approximation. The ideal gas equation of state $p\!=\!(\cP\!-\!\cV)
\rho T \!=\!\calR \rho T$ applies, where $\calR$ is the gas constant,
and $c_{\rm P}$ and $c_{\rm V}$ are the specific heats at constant
pressure and volume, respectively. The radiative flux is given by
$\FFF_{\rm rad}\!=\!-K\bm\nabla T,$ where $K$ is the radiative heat
conductivity, given by $K(\rho,T)\!=\!K_0 (\rho/\rho_0)^{-(a+1)}
(T/T_0)^{3-b}$. The choice $a\!=\!1$ and $b\!=\!-7/2$ corresponds to
Kramers opacity law \citep{Weiss_et_al_2004_Cox_Giuli_Principles},
which was first used in convection simulations by
\cite{Edwards_1990_MNRAS_242_224} and
\cite{Brandenburg_et_al_2000_Astrophysical_Convection_and_Dynamos_85}.
Turbulent SGS flux $\FFF_{\rm SGS}\!=\!-\chiSGS \rho T \bm\nabla s'$
applies to the entropy fluctuations
$s'(\xxx)\!=\!s(\xxx)\!-\!\mean{s}(z)$, where the overbar indicates
horizontal averaging and where $\chiSGS$ is a constant. $\FFF_{\rm
  SGS}$ has a negligible contribution to the net energy flux such that
$\mean{\FFF}_{\rm SGS} \approx 0$. The cooling function ${\cal C}$
consists of two parts: uniform cooling toward a constant temperature
above $\zcool$, and additional $\npatch$ local patches where the
cooling is applied already at a lower depth $\zpatch < \zcool$. The
cooling function is described in \Appendix{sec:cooling}. The advective
terms in \Equs{equ:dens}{equ:ent} contain a hyperdiffusive sixth-order
correction with a flow-dependent diffusion coefficient \citep[see
  Appendix~B of][]{DSB06}.

The computational domain extends between $z_{\rm bot}\!\leq
z\!\leq\!z_{\rm top}$ where $z_{\rm bot}/d\!=\!-0.45$ or $z_{\rm
  bot}/d\!=\!-0.95$ depending on the model, $z_{\rm top}/d\!=\!1.05$,
and the horizontal coordinates $x$ and $y$ run from $-2d$ to $2d$. The
initial stratification consists of three layers. The two lower layers
are polytropic with polytropic indices $n_1\!=\!3.25$ ($z_{\rm
  bot}/d\!\leq\!z/d\!\leq 0$) and $n_2\!=\!1.5$ ($0\!\leq\!z/d\!\leq
1$). The latter corresponds to a marginally stable isentropic
stratification. At $t\!=\!0$ the uppermost layer above $z/d\!=\!1$ is
isothermal, and convection ensues because the system is not in thermal
equilibrium. The velocity field is perturbed with small-scale Gaussian
noise with amplitude $10^{-5}\sqrt{dg}$. Horizontal boundaries are
periodic, and vertical boundaries are impenetrable and stress free. A
constant energy flux is imposed at the lower boundary by setting
$\pd_z T = -F_\tbot/K_\tbot$, where $F_{\rm bot}$ is a
fixed input flux and $K_\tbot\!=\!K(x,y,\zbot)$. A constant
temperature $T=T_{\rm top}$ is imposed on the upper vertical boundary.

The units of length, time, density, and entropy are given by
$[x]\!=\!d, [t]\!=\!\sqrt{d/g}, [\rho]\!=\!\rho_0, [s]\!=\!\cP$, where
$\rho_0$ is the initial value of density at $z=z_{\rm top}$. The
models are fully defined by choosing the values of $\nu$, $g$, $a$,
$b$, $K_0$, $\rho_0$, $T_0$, and the SGS Prandtl number
$\PraSGS\!=\!\nu/\chiSGS$, along with the parameters of the cooling
function. The quantity $\xi_0\!=\!\Hp^{\rm
  top}/d\!=\!\mathcal{R}T_{\rm top}/gd\!=\!  0.054$ is the initial
pressure scale height at the surface. The Prandtl number based on the
radiative heat conductivity is $\Pr(\xxx)\!=\!\nu/\chi(\xxx)$, where
$\chi(\xxx)\!=\!K(\xxx)/\cP \rho(\xxx)$. The dimensionless normalized
flux is given by $\Fn(\zbot)\!=\!\Fbot/\rho(\zbot) c_{\rm
  s}^3(\zbot)$, where $\rho(\zbot)$ and $c_{\rm s}(\zbot)$ are the
density and the sound speed, respectively, at $z/d\!=\!-0.45$ at
$t=0$. The Rayleigh number based on the energy flux is given by
$\RaF\!=\!gd^4 \Fbot/(\cP \rho T \nu\chi^2)$.

\begin{table}[t!]
\centering
\caption[]{Summary of the zones.}
  \label{tab:zones1}
       \vspace{-0.5cm}
      $$
          \begin{array}{cccccc}
          \hline
          \hline
          \noalign{\smallskip}
          \mbox{Zone} & \mFconv & \superad & \mbox{Label} & \mbox{Lower\ limit} & \mbox{Thickness} \\
          \hline
          \mbox{Buoyancy}  &    > 0    & >0 & \mbox{BZ} & \zbz & \dbz \\
          \hline
          \mbox{Deardorff} &    > 0    & <0 & \mbox{DZ} & \zdz & \ddz \\
          \hline
          \mbox{Overshoot} &    < 0    & <0 & \mbox{OZ} & \zoz & \doz \\
          \hline
          \mbox{Radiation} & \approx 0 & <0 & \mbox{RZ} &   -  &    -  \\
          \hline
          \end{array}
          $$ \tablefoot{The bottom of the overshoot zone is taken to
            be where $\mFkin$ falls below $10^{-2}\mFkin(z = \zdz)$
            similarly as in \cite{Kapyla_2024_AA_683_221}.}
\end{table}

\begin{table}[t!]
\centering
\caption[]{Summary of the runs.}
  \label{tab:runs1}
       \vspace{-0.5cm}
      $$
          \begin{array}{p{0.065\linewidth}ccccccccccc}
          \hline
          \hline
          \noalign{\smallskip}
          Run  & \npatch & \Rey  & \zbz  & \zdz  & \zoz  & \dbz  & \ddz  & \doz  & \dmix  & \fmix \\
          \hline
          A0    &   0  &  53  &   0.40  &  0.19  & -0.06  &  0.59  &  0.21  &  0.25  &  1.06  &  0.44 \\
          A10   &  10  &  47  &   0.46  &  0.17  & -0.20  &  0.54  &  0.28  &  0.37  &  1.19  &  0.55 \\
          A20   &  20  &  46  &   0.51  &  0.15  & -0.21  &  0.48  &  0.36  &  0.37  &  1.21  &  0.60 \\
          A50   &  50  &  38  &   0.58  &  0.10  & -0.42  &  0.40  &  0.48  &  0.52  &  1.40  &  0.71 \\
          A100  & 100  &  40  &   0.59  &  0.08  & -0.43  &  0.37  &  0.52  &  0.50  &  1.39  &  0.73 \\
          A200  & 200  &  53  &  0.35  &  0.10  & -0.21  &  0.61  &  0.25  &  0.31  &  1.17  &  0.48 \\
          \hline
          \end{array}
          $$ \tablefoot{Summary of the runs. $\PraSGS =1$ in all runs
            such that $\PeSGS=\Rey$. $\Rat = (4.4\ldots4.6) \cdot
            10^6$ in all runs, except in Run~A200 where $\Rat=5.5\cdot
            10^6$. The values of $\RaF$, and $\Fn$ from the initial
            state at $z=0$ are $4.9\cdot 10^4$ and $9.1\cdot 10^{-5}$,
            respectively. Runs~A0, A10, and A20 have $\zbot/d =-0.45$
            and use $288^3$ grid points, whereas Runs~A50, A100, and
            A200 have $\zbot/d =-0.95$ with a $288^2\times 384$ grid.}
\end{table}

\begin{figure*}
  \centering
  \includegraphics[width=0.5\textwidth,trim= 0 -0.5cm 0 0]{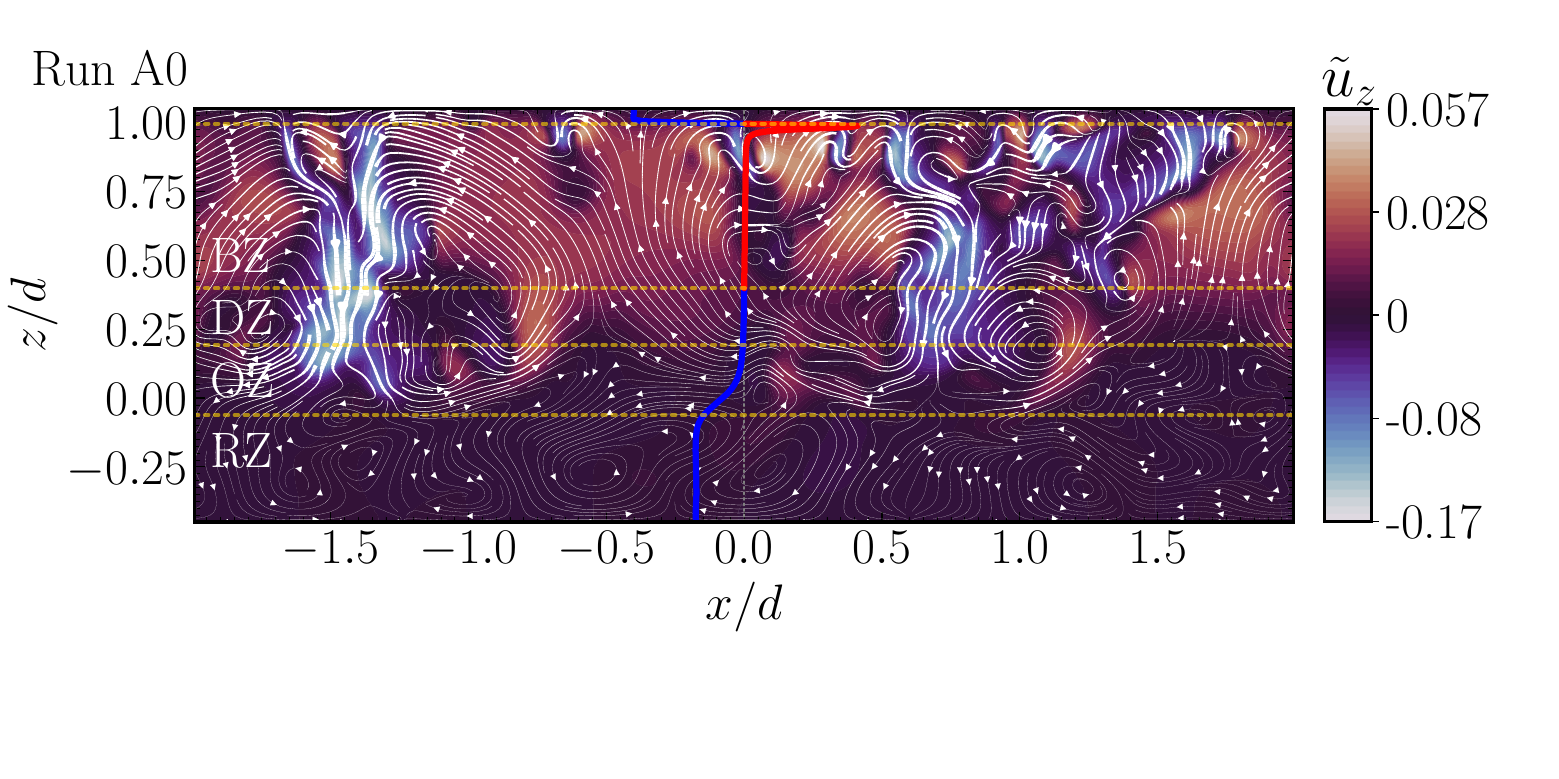}\includegraphics[width=0.5\textwidth]{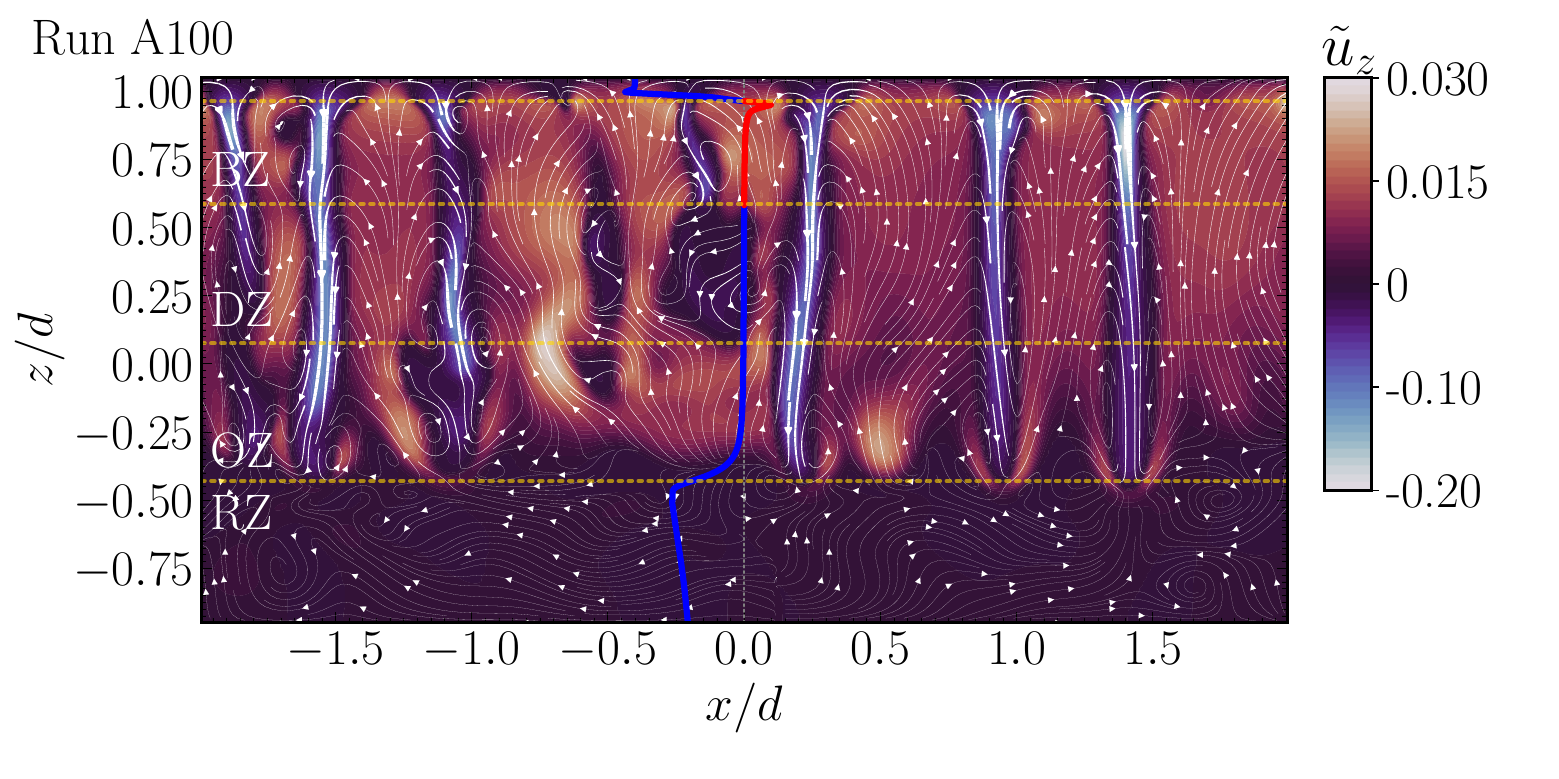}
  \vspace{-.8cm}
  \caption{Vertical slices showing the flows from Runs~A0 (left) and
    A100 (right). Note that the box is deeper in the latter case. The
    colour contours indicate the vertical velocity $u_z$. Tilde refers
    to normalization by $\sqrt{dg}$. The arrows map streamlines where
    the linewidth reflects the flow speed. The blue and red curve
    indicated the superadiabatic temperature gradient $\superad$ with
    negative (positive) values in blue (red). The boundaries of
    different zones defined in \Table{tab:zones1} are indicated by the
    corresponding labels and orange dotted lines.}
\label{fig:vertical_cuts}
\end{figure*}

The Reynolds and SGS P\'eclet numbers are given by $\Rey\!=\!
\urms/\nu k_1$, and $\Pe_{\rm SGS}\!=\!\PraSGS\Rey\!=\!\urms/\chiSGS
k_1$, where $\urms$ is the rms velocity averaged over the convectively
mixed layer and where $k_1\!=\!2\pi/d$. The total thermal diffusivity
is given by $\chieff(\xxx) = \chiSGS + \chi(\xxx)$. The turbulent
Rayleigh number $\Rat = \frac{gd^4}{\nu \mchieff}\left( -
\frac{1}{\cP}\frac{{\rm d}\mean{s}}{{\rm d}z} \right)$ is quoted from
$z/d=0.85$ in the statistically stationary state using the temporally
and horizontally averaged mean state denoted by the overbars.

\begin{figure*}
  \includegraphics[width=0.5\textwidth]{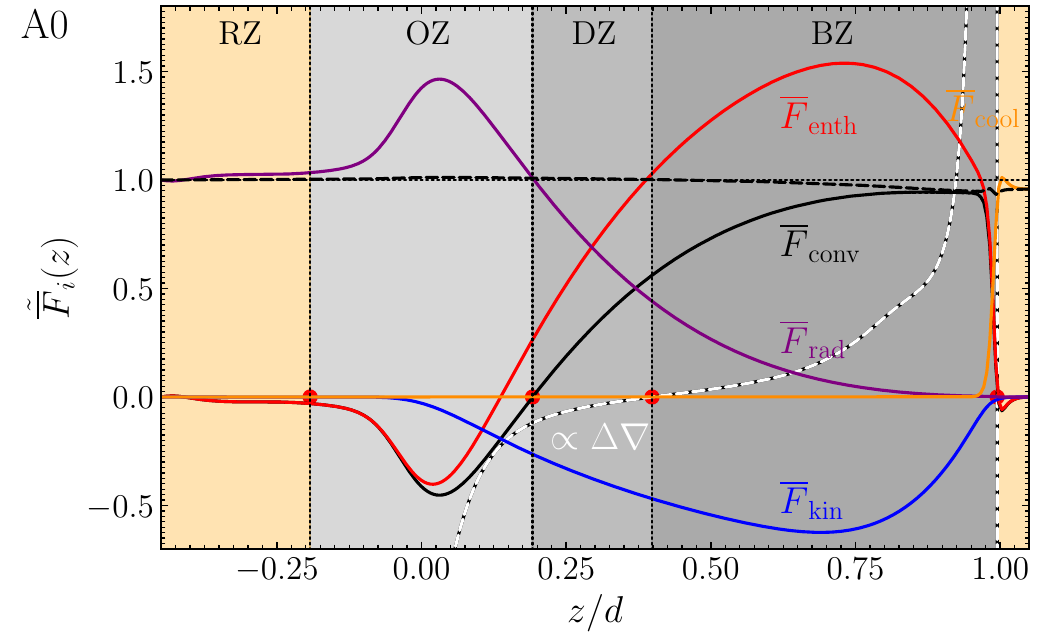}\includegraphics[width=0.5\textwidth]{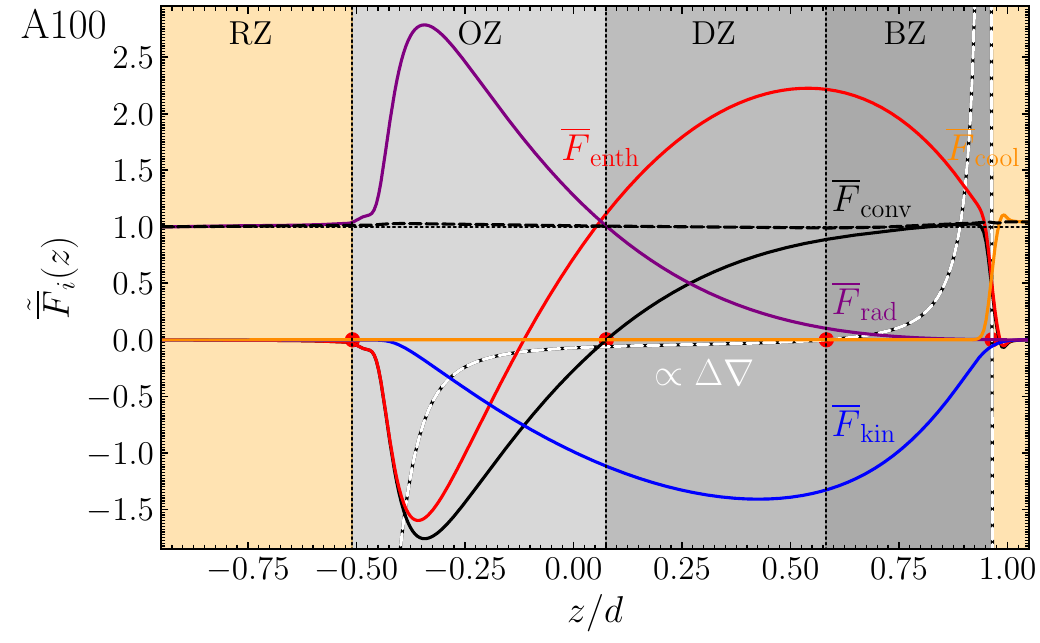}
  \caption{Horizontally averaged energy fluxes according to
    \Eqs{equ:fluxes1}{equ:fluxes2} along with $\mFconv$ from runs with
    $\npatch = 0$ (Run~A0) and $100$ (Run~A100). The dashed black and
    white line is proportional to $\superad$.}
\label{fig:plot_fluxes}
\end{figure*}

\section{Results}
\label{sec:results}

To study the structure of the convection zone, the dominant
contributions to the vertical energy flux need to be identified. These
are given by:
\begin{eqnarray}
  & \mFenth=\cP \mean{(\rho u_z)' T'},\ \ \mFkin=\onehalf \mean{\rho \uuu^2 u_z'}, & \label{equ:fluxes1} \\
  & \mFrad=-\mean{K}\pd_z \mean{T},\ \ \mFcool=-\int_{z_{\rm bot}}^{z} \mean{\mathcal{C}} {\rm d}z, & \label{equ:fluxes2}
\end{eqnarray}
and correspond to convective enthalpy and kinetic energy fluxes,
radiative flux and the surface cooling, respectfully. Here, the primes
denote fluctuations from horizontal averages. The total convected flux
is $\mFconv = \mFenth + \mFkin$
\citep{Cattaneo_et_al_1991_ApJ_370_282}. Furthermore, the
superadiabatic temperature gradient is $\superad=\nabla-\nabad$, where
$\nabla = \pd \ln T / \pd \ln p$. \Table{tab:zones1} summarizes the
different zones in the simulations based on the signs of $\mFconv$ and
$\superad$. The nomenclature used here is the same as in
\cite{Kapyla_et_al_2017_ApJL_845_23} apart from the fact that in the
latter $\mFenth$ instead of $\mFconv$ was used in the classification.

The main objectives of the study are to determine the effect of the
non-uniform entropy rain-inducing cooling on the structure of the
convection zone, convective energy transport and flows in the
convection zone. The runs and the most salient diagnostics are listed
in \Table{tab:runs1}.

\subsection{Convection zone structure and energy transport}

\Figu{fig:vertical_cuts} shows vertical cuts of the velocity field for
runs with $\npatch=0$ (Run~A0) and $100$ (A100). The former uses the
same set-up with uniform surface cooling as in several earlier studies
\citep[e.g.][]{Kapyla_2019_AA_631_122,Kapyla_2021_AA_655_78,
  Kapyla_2024_AA_683_221}, albeit with somewhat higher Reynolds
number. This run shows a typical pattern of moderately turbulent
stratified convection with cellular flows near the surface and
increasing scales of convective structures in deeper layers. The many
downflows near the surface around $z/d\approx 1$ merge to just a few
large downflows in the deep convection zone ($z/d\approx 0.2$). This
corresponds to a tree-like structure where the trunk is situated at
the base of the convection zone, and which is reminiscent of the
mixing length picture of convection. In Run~A100 with $\npatch=100$,
the structures related to large-scale cellular convection are absent
and the flow is dominated by deeply penetrating downflow plumes. These
downflows preserve their identity throughout the entire convection
zone forming a forest-like structure. Such configuration was envisaged
in the studies \cite{Spruit_1997_MEMSAI_68_397} and
\cite{Brandenburg_2016_ApJ_832_88}. The convectively mixed layer,
consisting of BZ, DZ, and OZ, is also significantly deeper in Run~A100
in comparison to Run~A0 due to the deeply penetrating
downflows. \Figu{fig:vertical_cuts} also shows that in the entropy
rain-dominated regime in Run~A100 the OZ and DZ are deeper, and the BZ
is shallower than in the case with uniform surface cooling (Run~A0).

\Figu{fig:plot_fluxes} shows the energy fluxes from
\Eqs{equ:fluxes1}{equ:fluxes2} from Runs~A0 and A100. Similar plots
for the other runs are shown in \Appendix{sec:fluxes}. A particularly
striking result is the increasing depth of the DZ and OZ when
$\npatch$ is increased. \Table{tab:runs1} lists the depths of
different zones for all of the current simulations. The depth of BZ
decreases, and the depth of DZ increases up to $\npatch=100$, whereas
OZ is already somewhat shallower in Run~A50 in comparison to
Run~A100. In the extreme case of $\npatch=200$ (Run~A200), the zone
structure is again approaching that of Run~A0 without cooling
patches. The most likely reason is plume merging and the fact that
even a small anisotropy in the placing of cooling patches leads to
horizontal pressure gradients that can drive large-scale circulation.

\begin{figure}
  \includegraphics[width=\columnwidth]{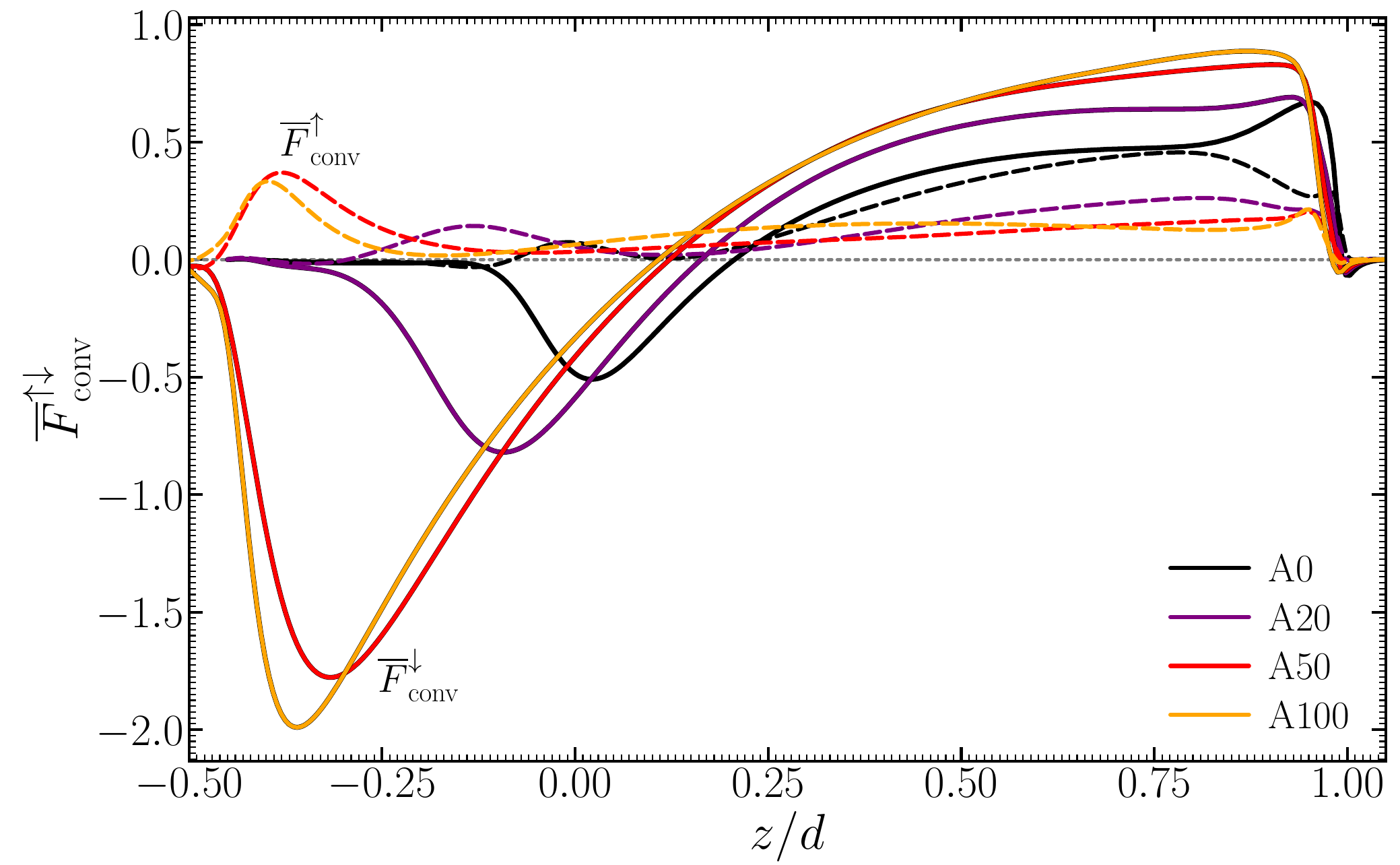}
  \caption{Net convected flux from downflows ($\mFconv^\downarrow$,
    solid lines) and upflows ($\mFconv^\uparrow$, dashed lines) from
    representative runs indicated by the legend.}
\label{fig:plot_flux_updown}
\end{figure}

\begin{figure}
  \includegraphics[width=\columnwidth]{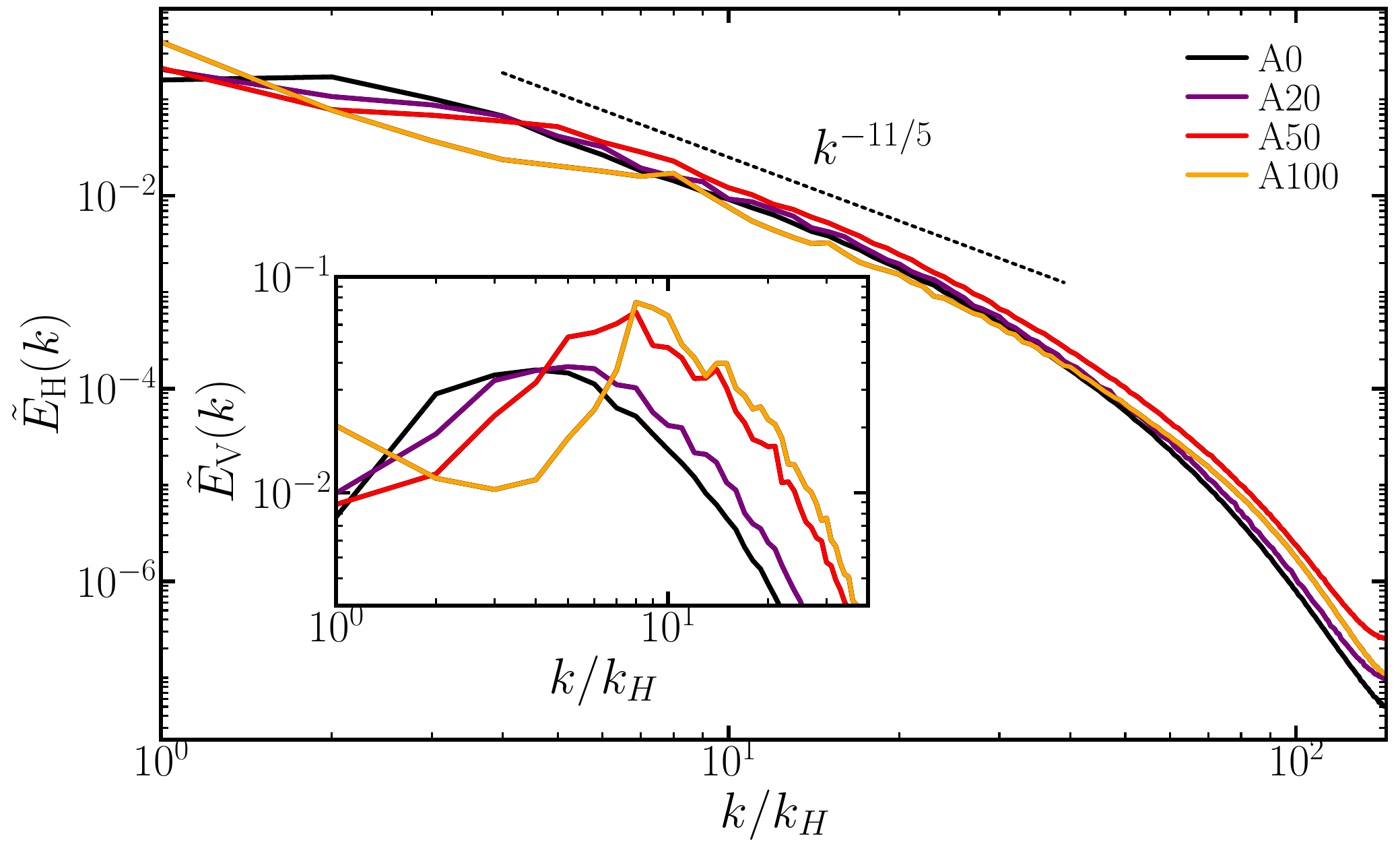}
  \caption{Power spectra of horizontal flows $E_{\rm H}(k)$ from the
    same runs as in \Fig{fig:plot_flux_updown} from
    $z/d=0.85$. Bolgiano-Obukhov $k^{-11/5}$ scaling is shown for
    reference. The inset shows the power spectrum of vertical flows
    $E_{\rm V}(k)$ for $k/\kH<40$ for the same runs. Tildes refer to
    normalization by the integrated spectrum for each case.}
\label{fig:plot_spectra_comp}
\end{figure}

The last two columns of \Table{tab:runs1} list the depth of the
convectively mixed layer $\dmix = \dbz+\ddz+\doz$ and the fraction of
$\dmix$ which is stably stratified according to the Schwarzschild
criterion, $\fmix=(\ddz+\doz)/\dmix$. Even in Run~A0 with no cooling
patches, over 40 per cent of the convectively mixed layer is stably
stratified. This fraction increases to 73 per cent in Run~A100, in
which case only about a quarter of the convectively mixed layer would
be considered convective in the classical mixing length paradigm.

The increasing depth of the subadiabatic regions is associated with
boosted magnitudes of the convective enthalpy flux $\mFenth$ and the
kinetic energy flux $\mFkin$. In Runs~A50 and A100 the maximum of
$\mFenth$ is nearly twice $\Fbot$ and the maxima of $|\mFkin|$ is
roughly $1.3\Fbot$; see
\Figas{fig:plot_fluxes}{fig:plot_fluxes_appendix}. The latter
indicates a significant increase in the energy transport due to
downflows. \Fig{fig:plot_flux_updown} shows the net convected flux
($\mFconv$) separately for downflows and upflows for representative
runs. In the fiducial Run~A0 the upflows and downflows transport
roughly equal amounts of flux in the bulk of the CZ between
$0.2\lesssim z/d\lesssim 0.8$. When $\npatch$ increases, the share of
the net energy flux carried by the downflows increases such that for
$\npatch=50$ (Run~A50) and $\npatch=100$ (Run~A100) the upflows
contribute only about $0.2\Fbot$ whereas the downflows transport the
majority of the flux with $0.8\Fbot$. For $\npatch=200$ the flux
balance resembles more the case with no plumes because large-scale
convection ensues in this case; see \Fig{fig:plot_fluxes_appendix}.

\subsection{Spatial scale of convection}

The power spectrum of horizontal velocities $E_{\rm H}(k)$ at
$z/d=0.85$ from representative runs are shown in
\Fig{fig:plot_spectra_comp}. The maximum of $E_{\rm H}(k)$ is at the
largest possible scale irrespective of the boundary forcing. This is
likely due to the aforementioned sensitivity to the placement of the
cooling patches leading to horizontal pressure gradients that drive
flows possibly on very large scales. This is especially an issue in
the cases studied here where the cooling patches are stationary. On
intermediate scales ($5\lesssim k/\kH \lesssim 40$) the power spectra
for the horizontal flows is close to the Bolgiano-Obukhov $k^{-11/5}$
scaling similarly to \cite{Kapyla_2021_AA_655_78} and
\cite{Warnecke_et__al_2024arXiv240608967W}. The peak of the power
spectrum of vertical flows ($E_{\rm V}(k)$) shifts toward smaller
scales (higher $k$) as $\npatch$ increases, reflecting the dominance
of the plumes for vertical flows. In \Fig{fig:plot_spectra_comp}
normalization is chosen such that differences between runs appear in
the shape of the spectra. An alternative presentation highlighting the
absolute differences between runs is shown in
\Appendix{sec:power_aniso}.

\section{Conclusions}
\label{sec:conclusions}

The current proof-of-concept simulations show that strong entropy rain
can radically change the convective flows and the ensuing energy
transport as well as the thermal structure of the convection zone. In
the entropy rain-dominated regime convection is dominated by downflows
plumes penetrating the whole convection zone rendering the majority of
the convectively mixed layer weakly stably stratified along the
pioneering ideas of \cite{Spruit_1997_MEMSAI_68_397} and
\cite{Brandenburg_2016_ApJ_832_88}. The idea of entropy rain-dominated
convection in the Sun coincides with the recent studies of solar
Rossby waves that suggest weak subadiabaticity of deep convection zone
\citep{Bekki_2024_AA_682_39} and possible fast small-scale downflows
at supergranular scales \citep{Hanson_et_al_2024_NatAs_8_1088}.

The current simulations employ a very simple setup and probe a very
modest portion of the presently accessible parameter
space. Furthermore, rotation and magnetic fields, that are dynamically
important in stellar interiors, are neglected. Although a recent study
\citep{Kapyla_2024_AA_683_221} suggests that convection in the Sun is
not strongly rotationally constrained anywhere, its effect is perhaps
more subtle in the entropy rain-dominated regime because the slow and
spatially larger upflows are much more affected by rotation than the
fast and small downflows. The increasing anisotropy between upflows
and downflows can also have implications also for large-scale magnetic
field generation due to convection. The nonlinear backreaction of
small-scale magnetic fields on convection
\citep{Hotta_et_al_2022_ApJ_933_199}. is another aspect that cannot in
general be neglected. These effects will be explored further in future
studies.

\begin{acknowledgements}
  I acknowledge the stimulating discussions with participants of the
  Nordita Scientific Program on ”Stellar Convection: Modelling, Theory
  and Observations”, in August and September 2024 in Stockholm. The
  simulations were performed using the resources granted by the Gauss
  Center for Supercomputing for the Large-Scale computing project
  ``Cracking the Convective Conundrum'' in the Leibniz Supercomputing
  Centre's SuperMUC-NG supercomputer in Garching, Germany.
\end{acknowledgements}

\bibliographystyle{aa}
\bibliography{paper}

\appendix

\section{Cooling function}
\label{sec:cooling}

The cooling near the surface is described be two parts
\begin{eqnarray}
\mathcal{C} = \mathcal{C}_1 + \mathcal{C}_2,
\end{eqnarray}
where
\begin{eqnarray}
\mathcal{C}_1 = \rho\cP \frac{T_{\rm cool} - T}{\taucoolo} f_{\rm cool}(z,\zcool),
\label{equ:cool}
\end{eqnarray}
where $\taucoolo = 0.2 \sqrt{d/g}$ is a cooling time, $T=e/c_{\rm V}$
is the temperature, $e$ is the internal energy, and $T_{\rm
  cool}=T_{\rm top}$ is a reference temperature corresponding to the
fixed value at the top boundary. Furthermore,
\begin{eqnarray}
f_{\rm cool}(z,\zcool) = \onehalf \left[ 1 + \tanh \left({\frac{z-\zcool}{\dcool}}\right)\right],
\end{eqnarray}
where $\zcool/d = 1$ and $\dcool=0.01d$ for $\mathcal{C}_1$.

In addition, $\npatch$ localized cooling patches are introduced via
$\mathcal{C}_2$. Each patch follows:
\begin{eqnarray}
  \mathcal{C}_2 &=& \rho\cP \frac{T_{\rm cool} - T}{\taucoolt} \fcool(z,\zpatch) \times \nonumber \\
   && \hspace{2cm} \exp \left( -\onehalf \frac{(x-\xpatch)^2 + (y-\ypatch)^2 }{\dpatch^2} \right),
\end{eqnarray}
where $\taucoolt = 0.067 \sqrt{d/g}$, $\fcool$ is the same profile as
above with $\zpatch/d=0.96$, and where the width of the patches is
$\dpatch=0.05d$ Since $\zpatch < \zcool$, the localized cooling
corresponds to cool fingers penetrating the upper part of the
convection zone. The number of patches ($\npatch$) is another input to
the models. In the present study the positions ($\xpatch,\ypatch$) of
the patches are fixed in time and chosen such that $\sum_i^{\npatch}
\xpatch^{(i)} = 0$ and $\sum_i^{\npatch} \ypatch^{(i)} = 0$.

\section{Energy fluxes}
\label{sec:fluxes}

\begin{figure*}
  \includegraphics[width=0.5\textwidth]{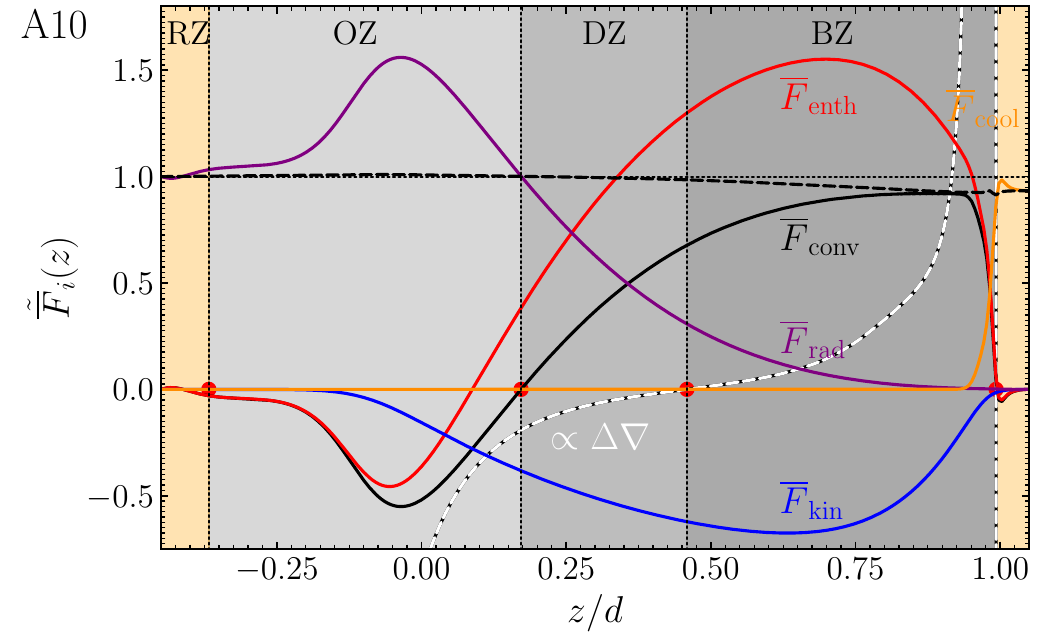}\includegraphics[width=0.5\textwidth]{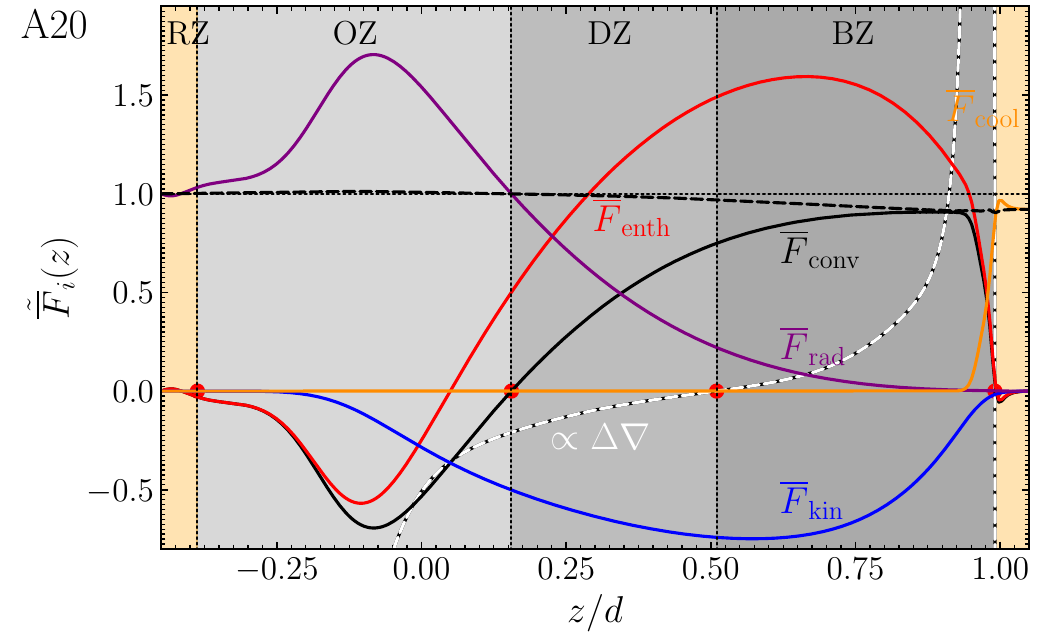}
  \includegraphics[width=0.5\textwidth]{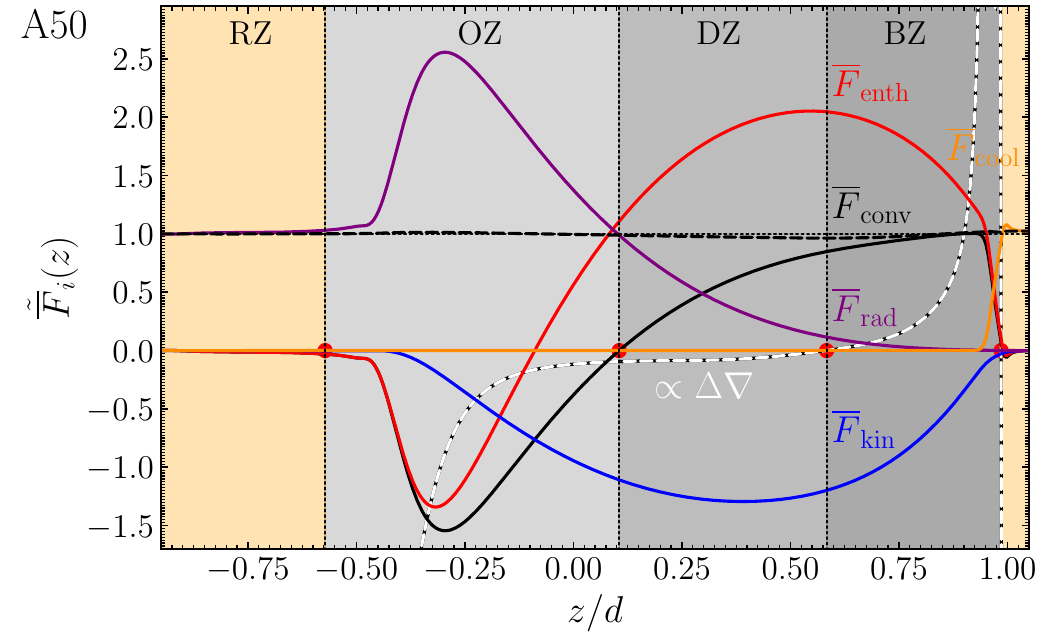}\includegraphics[width=0.5\textwidth]{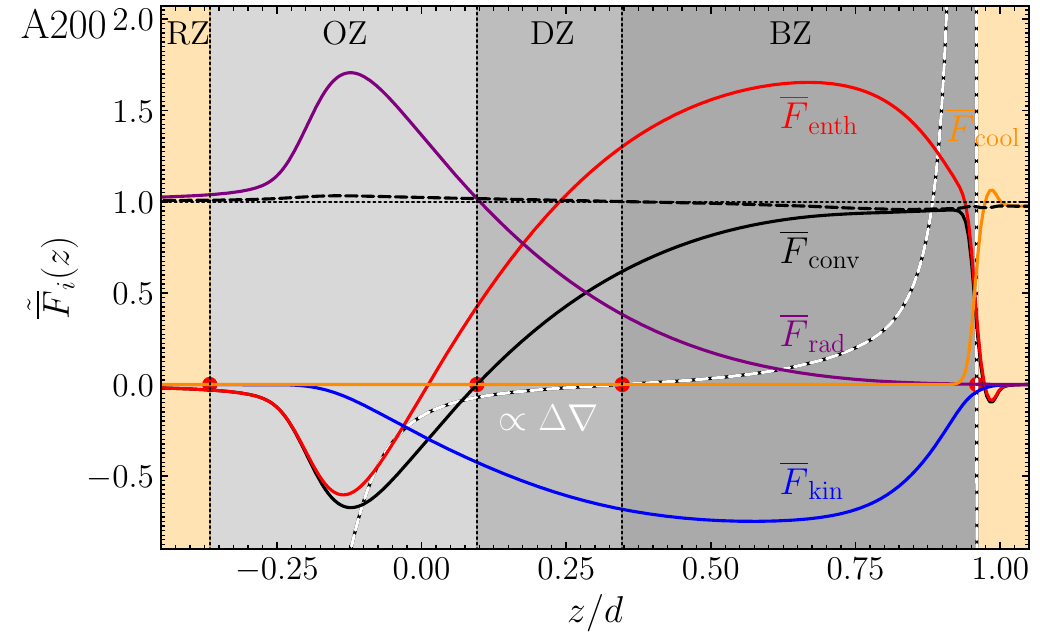}
  \caption{Horizontally averaged energy fluxes according to
    \Eqs{equ:fluxes1}{equ:fluxes2} along with $\mFconv$ from runs with
    $n_{\rm plume} = 10$, $20$, $50$, and $200$.}
\label{fig:plot_fluxes_appendix}
\end{figure*}

\Figu{fig:plot_fluxes_appendix} shows the energy fluxes from Runs~A10,
A20, A50, and A200.

\begin{figure}
  \includegraphics[width=\columnwidth]{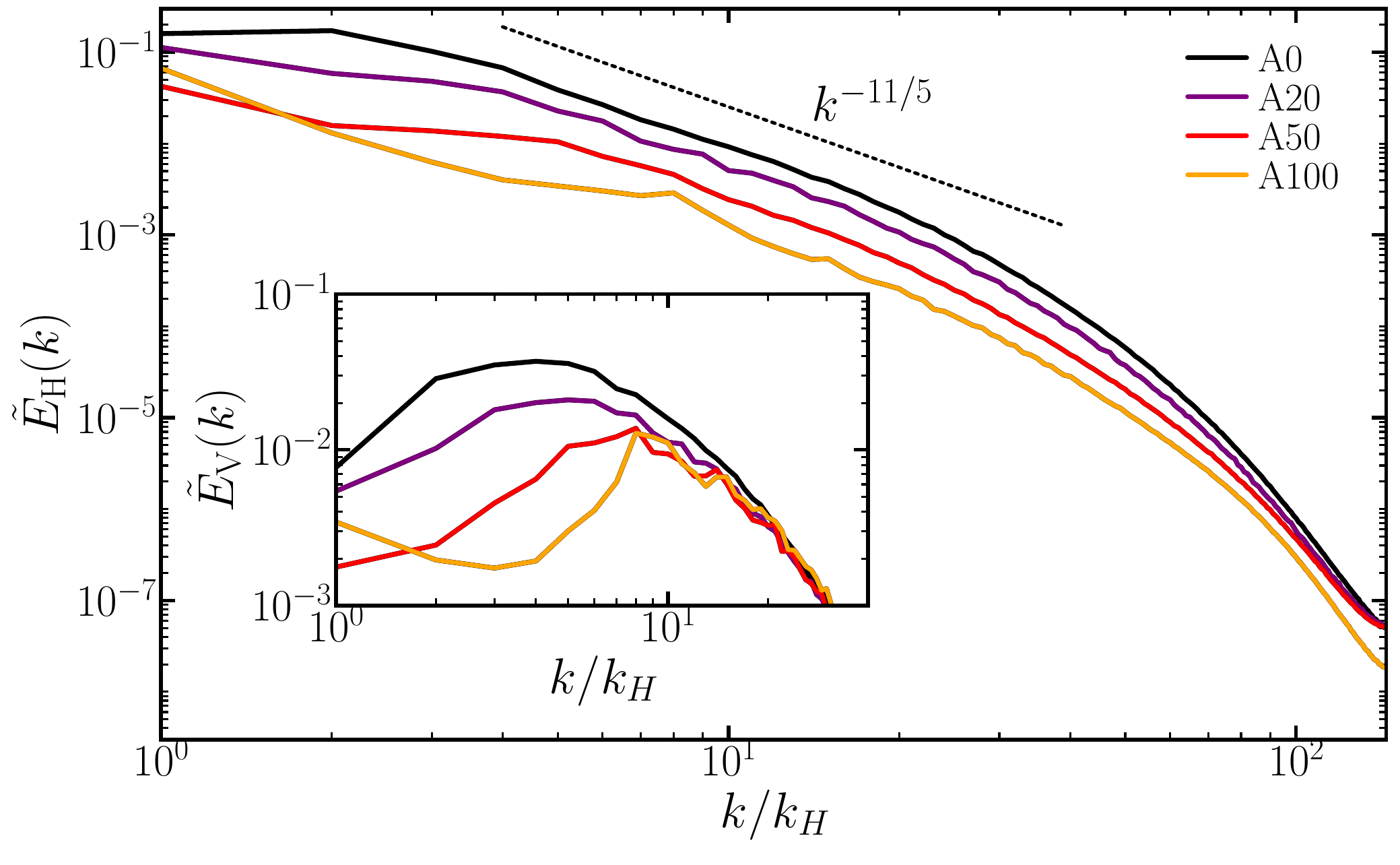}
  \caption{Same as \Fig{fig:plot_spectra_comp} but normalization with
    respect to the integrated spectra of Run~A0.}
\label{fig:plot_spectra_comp_norm}
\end{figure}

\begin{figure}
  \includegraphics[width=\columnwidth]{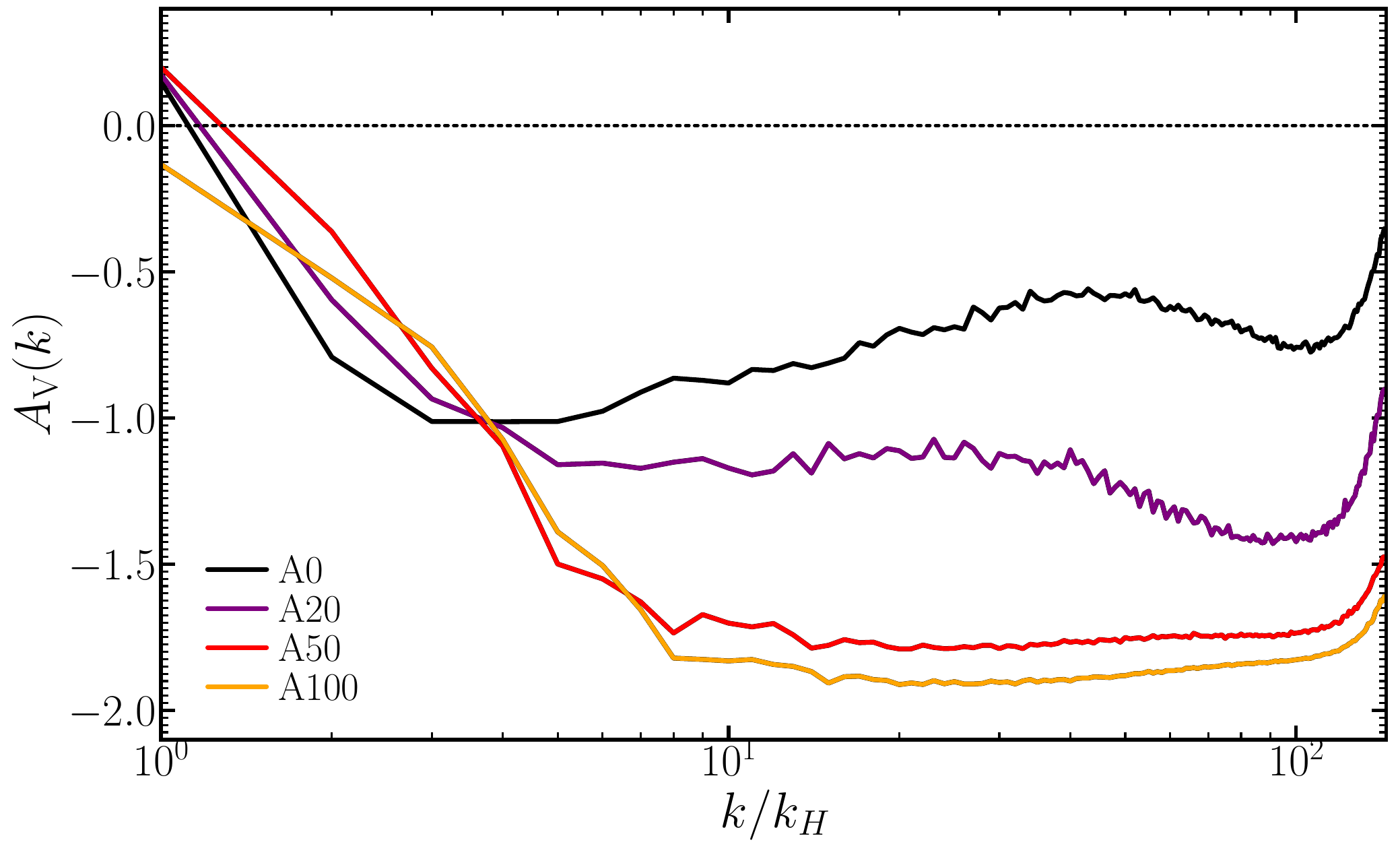}
  \caption{Anisotropy parameter $\AAV(k)$ from the same runs as in
    \Fig{fig:plot_spectra_comp}.}
\label{fig:plot_Av_comp}
\end{figure}

\begin{figure}
  \includegraphics[width=\columnwidth]{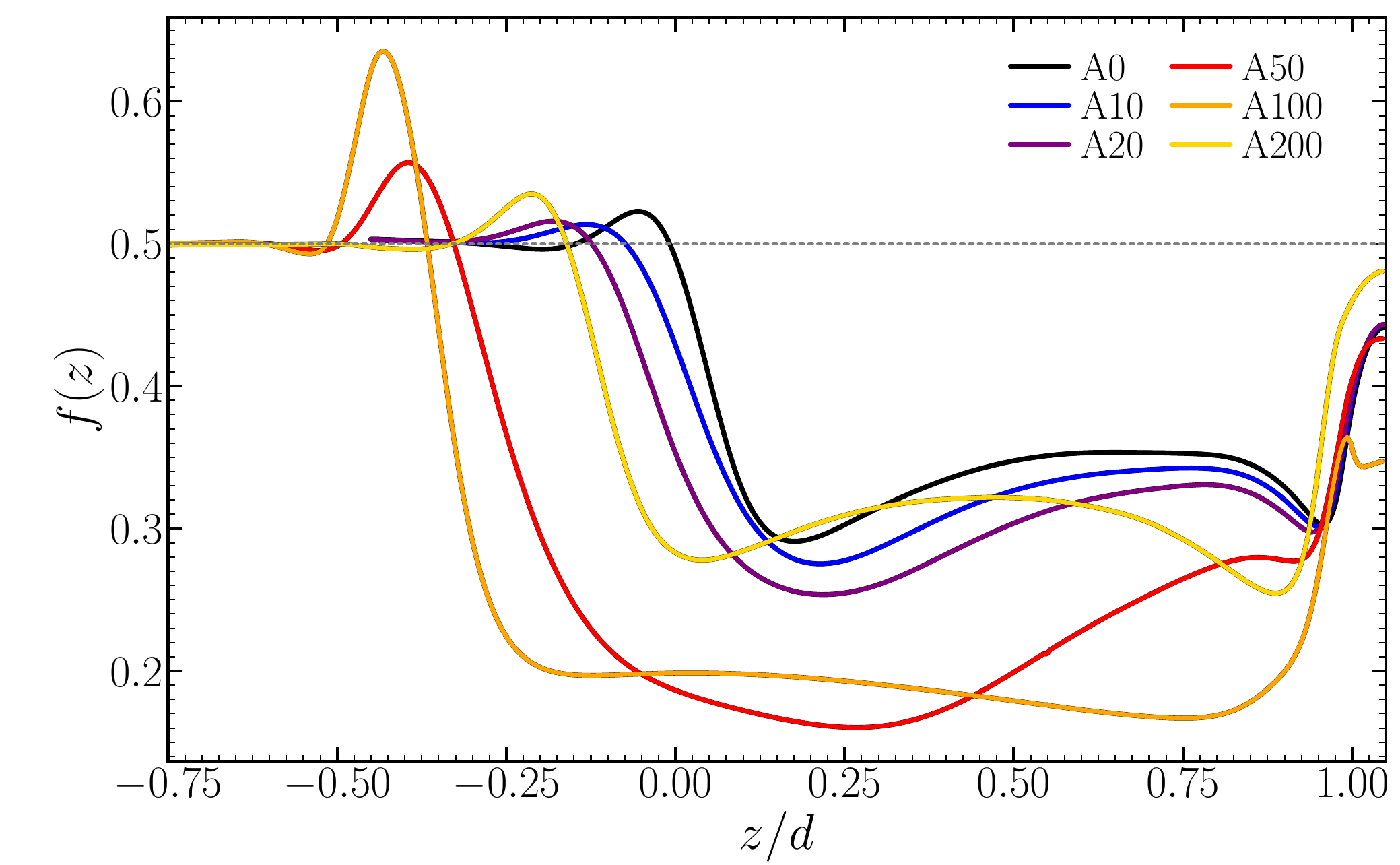}
  \caption{Filling factor of downflows as a function of depth from all
    of the current runs.}
\label{fig:plot_ff}
\end{figure}

\section{Power spectra and anisotropy}
\label{sec:power_aniso}

The horizontal and vertical power spectra are defined such that
\begin{eqnarray}
\int_0^{\kmax} E_{\rm V}(k,z) dk &=& \onehalf \mean{u_z^2}(z), \\
\int_0^{\kmax} E_{\rm H}(k,z) dk &=& \onehalf [\mean{u_x^2}(z) + \mean{u_y^2}(z)].
\end{eqnarray}
\Figu{fig:plot_spectra_comp_norm} shows the vertical and horizontal
spectra from the same runs as in \Fig{fig:plot_spectra_comp} but
normalized with respect to the integrated spectra in Run~A0. Both the
vertical and horizontal velocities decrease with $\npatch$. The
anisotropy of the flow can be characterized with the vertical spectral
anisotropy parameter \citep{Kapyla_2019_AN_340_744}:
\begin{eqnarray}
  \AAV(k,z) = \frac{\ESH(k,z)-2\ESV(k,z)}{\ESK(k,z)},
\end{eqnarray}
where $\ESK(k,z) = \ESH(k,z) + \ESV$(k,z). The anisotropy parameter is
shown in \Fig{fig:plot_Av_comp} from $z/d = 0.49$ which is near the
middle of the convectively mixed layer. While the anisotropy at large
scales is weak irrespective of $\npatch$, it is significantly enhanced
at intermediate and small scales ($k/\kH \gtrsim 7$). The increased
anisotropy parallels the increasing dominance downflows in the energy
transport.

A further diagnostic of the asymmetry of convective flows is the
filling factor of downflows $f^\downarrow$, defined via
\begin{eqnarray}
\mean{u}_z = f^\downarrow \mean{u}_z^{\downarrow} + (1 - f^\downarrow)
\mean{u}_z^{\uparrow},
\end{eqnarray}
where $\mean{u}_z^\downarrow$ and $\mean{u}_z^\uparrow$ are the
averaged downflows and upflows, respectively. \Figu{fig:plot_ff} shows
that the filling factor decreases monotonically everywhere when more
downflow plumes are added until $\npatch = 50$. This can be understood
such that in cases with many plumes, cellular convection is excited
only near the surface whereas deeper down the average stratification
is stably stratified and plumes are almost the only downflow
structures. For $\npatch = 100$ (Run~A100) the spatial profile of
$f^\downarrow$ changes although it is still on average lower than for
the cases with fewer plumes. For $\npatch=200$ (Run~A200) the filling
factor is again similar to the cases with few or no imposed
plumes. This is explained by merging of plumes already near the
surface into larger downflows and the development of large-scale
overturning convection in that case.

\end{document}

%% file: macros.tex
\usepackage{xcolor}

\newcommand{\uuu}{{\bm u}}

\newcommand{\xxx}{{\bm x}}

\newcommand{\FFF}{{\bm F}}

\newcommand{\SSt}{\bm{\mathsf{S}}}
\newcommand{\SStij}{\mathsf{S}_{ij}}

\newcommand{\Eqs}[2]{Equations~(\ref{#1}) and (\ref{#2})} 
\newcommand{\Equs}[2]{Equations~(\ref{#1}) to (\ref{#2})}

\newcommand{\EQ}{\begin{equation}}
\newcommand{\EN}{\end{equation}}
\newcommand{\EQA}{\begin{eqnarray}}
\newcommand{\ENA}{\end{eqnarray}}

\newcommand{\pd}{\partial}

\newcommand{\mean}[1]{\overline{#1}}

\newcommand{\cP}{c_{\rm P}}
\newcommand{\cV}{c_{\rm V}}

\newcommand{\urms}{u_{\rm rms}}

\newcommand{\AAV}{A_{\rm V}}

\newcommand{\kmax}{k_{\rm max}}

\newcommand{\kH}{k_{\rm H}}

\newcommand{\chiSGS}{\chi_{\rm SGS}}

\newcommand{\chieff}{\chi_{\rm eff}}
\newcommand{\mchieff}{\overline{\chi}_{\rm eff}}

\newcommand{\Hp}{H_{\rm p}}

\newcommand{\Pe}{{\rm Pe}}

\newcommand{\PeSGS}{{\rm Pe}_{\rm SGS}}

\newcommand{\PraSGS}{{\rm Pr}_{\rm SGS}}

\newcommand{\Rat}{{\rm Ra}_{\rm t}}
\newcommand{\RaF}{{\rm Ra}_{\rm F}}

\newcommand{\Rey}{{\rm Re}}

\newcommand{\taucoolo}{\tau_{\rm cool1}}
\newcommand{\taucoolt}{\tau_{\rm cool2}}

{}

\newcommand{\calR}{{\cal R}}

\newcommand{\Fbot}{F_{\rm bot}}

\newcommand{\Fn}{\mathscr{F}_{\rm n}}

\newcommand{\mFenth}{\mean{F}_{\rm enth}}
\newcommand{\mFconv}{\mean{F}_{\rm conv}}
\newcommand{\mFrad}{\mean{F}_{\rm rad}}
\newcommand{\mFkin}{\mean{F}_{\rm kin}}

\newcommand{\mFcool}{\mean{F}_{\rm cool}}

\newcommand{\ESK}{E_{\rm K}}
\newcommand{\ESH}{E_{\rm H}}
\newcommand{\ESV}{E_{\rm V}}
\newcommand{\nabad}{\nabla_{\rm ad}}

\newcommand{\superad}{\Delta\nabla}

\newcommand{\tbot}{{\rm bot}}

\newcommand{\dpatch}{d_{\rm patch}}
\newcommand{\xpatch}{x_{\rm patch}}
\newcommand{\ypatch}{y_{\rm patch}}

\newcommand{\zbot}{z_{\rm bot}}
\newcommand{\zbz}{z_{\rm BZ}}

\newcommand{\zdz}{z_{\rm DZ}}

\newcommand{\zoz}{z_{\rm OZ}}

\newcommand{\zcool}{z_{\rm cool}}

\newcommand{\zpatch}{z_{\rm patch}}
\newcommand{\dbz}{d_{\rm BZ}}

\newcommand{\ddz}{d_{\rm DZ}}

\newcommand{\dmix}{d_{\rm mix}}

\newcommand{\doz}{d_{\rm OZ}}

\newcommand{\dcool}{d_{\rm cool}}
\newcommand{\fcool}{f_{\rm cool}}
\newcommand{\fmix}{f_{\rm mix}}
\newcommand{\npatch}{n_{\rm patch}}

\def\onethird{{\textstyle{1\over3}}}

\def\onehalf{{\textstyle{1\over2}}}

\newcommand{\Figas}[2]{Figs.~\ref{#1} and \ref{#2}}

\newcommand{\Fig}[1]{Figure~\ref{#1}} 
\newcommand{\Figu}[1]{Figure~\ref{#1}}

\newcommand{\Table}[1]{Table~\ref{#1}}

\newcommand{\Appendix}[1]{Appendix~\ref{#1}}

\definecolor{ForestGreen}{RGB}{34,139,34}
\definecolor{AGray}{rgb}{.4,.4,.4}
\definecolor{LightYellow}{rgb}{1.,1.,.8}
\definecolor{LightCyan}{rgb}{0.88,1,1}